# Seebeck Nanoantennas for Solar Energy Harvesting

E. Briones[1,*], J. Briones[2], A. Cuadrado[3], J. C. Martinez-Anton[3], S. McMurtry[4], M. Hehn[4], F. Montaigne[4], J. Alda[3] and F. J. Gonzalez[1]

[1] CIACyT, Universidad Autonoma de San Luis Potosi, San Luis Potosi, 78210 SLP, Mexico

[2] Department of Mathematics and Physics, ITESO, Jesuit University of Guadalajara, 45604, Mexico

[3] Faculty of Optics and Optometry, Universidad Complutense de Madrid, 28037, Madrid, Spain

[4] Institut Jean Lamour, CNRS, Université de Lorraine, F-54506 Vandoeuvre Les Nancy, France

We propose a mid-infrared device based on thermocouple optical antennas for light sensing and energy harvesting applications. We numerically demonstrate that antennas are able to generate low-power dc signals by beneficing of the thermoelectric properties of the metals that constitute them. We theoretically evaluate the optical-to-electrical conversion efficiency for harvesting applications and finally discuss strategies to increase its performance. Thermocouple optical antennas therefore open the route toward the design of photovoltaic devices.

---

[*] Author to whom correspondence should be addressed. Electronic mail: edgar.briones@uaslp.mx

Over the last decade, the idea of using resonant nanoantennas for solar energy harvesting has gained considerable attention [1,2], as they introduce a means to capture the optical energy of free-propagating waves and to localize it into small volumes, providing thus an enabling technology for energy gathering in the far-infrared region [3,4]. These nanostructures take advantage of the wave nature of the electromagnetic radiation in order to transfer the optical energy to localized resonant currents [5,6], which are subsequently exploited to recover or sense the confined energy [7-9], opening thereby a route for the engineering of solar devices.

A successful incorporation of nanoantennas into photovoltaic technology relies on the implementation of an efficient retrieving mechanism, currently non-existent. In this context, nanoantennas coupled to high-speed rectifiers (known as "rectennas"), based on tunnel barriers, have been extensively explored during the last years as optical harvesting devices [10-14]. In spite of the high theoretical efficiency they can reach [15-17], rectennas exhibit low efficiencies due to the poor performance of the current rectifiers at optical frequencies [18]. As we wait for an efficient rectifier to be developed, other harvesting mechanisms must be explored [19].

In this letter, we explore a device which combines the use of nanoantennas (to confine the optical energy) with the thermoelectric properties of their metallic interfaces [20-22] (in order to recover such energy). The proposed devices consist of metallic thermocouples shaped as nano-antennas (Seebeck nanoantennas) sized to resonate to mid-infrared wavelengths. We show by means of simulations that these devices work by exploiting the temperature gradient caused by the resonant currents in the structures, which in turn generate a dc voltage $V_{OC}$ by Seebeck effect at their "open ends" [23,24]; hence defining a mechanism to harvest the optical energy.

We investigate the conversion efficiency of Seebeck nanoantennas from both theoretical and numerical simulations approaches, evaluating their potential as power micro-generators that retrieve the infrared energy. We analyze two types of structures, a square and an Archimedean spiral antennas (shown in Fig. 1), widely used for broadband applications at microwave and GHz frequencies, here scaled-down to infrared wavelengths [25-27]. The analysis is performed by using the commercial software COMSOL Multiphysics (ver3.5a) based on the finite-element method. This commercial package includes a platform where both the electromagnetic and thermal domains are fully integrated, as required for the present analysis.

We choose the so-called self-similarly spiral nanoantennas as convenient systems for solar applications since their electrical impedance behavior is ideally frequency-independent that leads to a broadband optical absorption [28]. These types of structures respond to any linearly polarized mid-infrared wave but exhibit a better performance for the right-handed circularly polarized light (RHCP) [26], confining the electric field $E$ at its center. However, left-handed circularly polarized light (LHCP) spirals should be also included into a single device in order to fully benefit of the unpolarized sunlight.

Thin-film thermocouples are then conformed from spiral nanoantennas by building up their arms with dissimilar metals. The chosen power generators have one Ti-Ni interface placed at the center of the structures, what indeed optimizes the harvesting of thermal energy (as Joule heating is stronger there). On the other hand, materials with a considerable difference in their Seebeck coefficients $S_A$ and $S_B$ (property that determines the performance of materials to generate voltage from heat gradients) should be considered in order to increase the response of the devices, given by [23,24]:

$$V_{OC} = (S_A - S_B) \Delta T \qquad (1)$$

where $\Delta T$ is the temperature difference between the center (hot spot) and the open ends of spirals (cold spots). In this analysis we have chosen nickel and titanium as building materials, as they show a considerable difference in their Seebeck coefficients ($S_{Ni}$ = -19.5 µV/K and $S_{Ti}$ = 7.19 µV/K [29]) and combine their relatively low thermal conductivities to reach adequate temperature gradients along the structures ($\kappa_{Ni}$ = 90 W/m K and $\kappa_{Ti}$ = 21.9 W/m K).

The Seebeck nanoantennas lie on the top of a silicon-oxide half-space which acts as a good thermal insulator allowing the structures to reach stronger temperature gradients. Actually, a $SiO_2$ layer of 1.2 µm would be enough to obtain an efficient thermal isolation. Since the dielectric substrate also modifies the optical resonances of the antennas to longer wavelengths [30,31]), the lengths were respectively adjusted to 40 µm and 25.8 µm (with a cross section of 100 nm width × 200 nm thickness that optimizes the optical absorption [32]), to appropriately operate in the mid-infrared region at 10.6 µm. The optical model was built by using the experimentally reported dielectric function of metals [33] and a monochromatic plane wave is used for normal far-field illumination. The irradiance $S$ of the plane wave was adjusted to 117 W/cm$^2$ for each single frequency in order to find the response of the devices. This irradiance is usually employed to experimentally test similar antenna-coupled devices [9,26].

Once the Seebeck nanoantennas have been designed, the first step involves determining the temperature distribution caused in devices by the incident light. For this purpose we perform a series of thermal simulations in which the antennas are considered as being the main source of heat, where heating is generated by the ohmic losses due to the resonant currents [34,35]. The heat density ($q$ [W/m$^3$]) along the spiral antennas is found using the Joule effect expression $q = \boldsymbol{J} \bullet \boldsymbol{E}$, where $\boldsymbol{E}$ denotes the electric field distribution and $\boldsymbol{J}$ the resonant current. This can be expressed in terms of the electric field $\boldsymbol{E}$ and the permittivity function of metals $\varepsilon_m$ as $q = \omega\, \varepsilon_0\, \text{Im}(\varepsilon_m)\, |\boldsymbol{E}(r)|^2$, where $\omega$ is the angular frequency of the electromagnetic wave. The heat source density $q$ is then used to numerically solve the conduction heat-transfer equation in order to find the steady-state temperature distribution $T(r)$ of the thermal system, inside and outside the nanoantennas (by imposing that substrate is thermalized at room temperature).

The results of the finite-element electromagnetic and thermal simulations are finally employed to evaluate the harvesting efficiency of the Seebeck nanoantennas. The optical-to-electrical conversion efficiency $\eta_e$ is here evaluated as the ratio of the power generated by thermocouples $P_{DC}$ to the optical power collected by them $P_{rec}$, expressed as:

$$\eta_e = \frac{P_{DC}}{P_{rec}} \qquad (2)$$

where the optical power $P_{rec} = A_{eff} \times S$ is found by using the reported collection area of the antennas $A_{eff}$ (12.5 µm$^2$ and 4 µm$^2$, for the square and Archimedean spiral respectively [26]) and the irradiance of the incidence beam $S$ set for the simulations.

In order to determine the power $P_{DC}$ that the thin-film thermocouples can generate, we consider the antenna circuit closed through a load resistance $R_L$ which causes a dc current flow $I$ and power exchange between both elements (Fig. 2). This load resistance is desired to act as a room-temperature bath that will increase the temperature gradient along the structures. The thermocouple power is then expressed by means of the relationship [23,36-40]:

$$P_{DC} = (S_A - S_B) \Delta T\ I - R_i\ I^2 \qquad (3)$$

where $R_i$ refers to the internal resistance of thermocouples (which generates dc ohmic losses moving the heat to the cold side of structures). On the other hand, a maximum value of power can be supplied from thermocouples to the load when both elements are electrically matched, given rise to power transfer [38]:

$$P_{DC}^{MAX} = \frac{(S_A - S_B)^2 \Delta T^2}{4\ R_i} = \frac{V_{OC}^2}{4\ R_i} \qquad (4).$$

Here, we estimate the overall conversion efficiency $\eta_e$ by considering the system of best matching; using thus the maximum transferred power between the antennas and the load to evaluate $\eta_e$. With the purpose of evaluating the maximum $P_{DC}$, the internal resistance $R_i$ of the devices is ideally considered as the dc ohmic resistance of the thermocouples arms [23,40], evaluated as 490 Ω and 310 Ω for the square and Archimedean spiral thermocouples, respectively.

Fig. 3(a) shows a series of temperature maps for the spirals thermocouples with different polarization states (RHCP and LHCP) for the particular case of an incident 10.6 μm plane wave. The temperature distribution depends on the polarization state of radiation since the spiral structures are sensitive to it (RHCP for the base case of coupling), unveiling further the amplitude of the resonant current along the arms. For all the polarization states a slight difference of the heating between the nickel and titanium arms is due to the difference in its thermal conductivity.

The temperature profiles for all these cases are shown in Fig. 3(b). In particular, for the case of best optical matching, the Ti-Ni junction shows a clear behavior as hot spot (open circles) and extremes as cold spots (filled circles), reaching temperature differences around $\Delta T \approx 215$ mK and 340 mK and generating low-voltage signals of 5.74 µV and 9.08 µV for the square and Archimedean spirals, respectively .

In Fig. 4(a) and 4(b), we plot the open voltage and the power that antennas are able to generate and transmit to the load resistance. These two values were numerically obtained as a function of the arriving wavelength, covering the mid-infrared region (from 3 µm to 50 µm), keeping the optical energy flux-density constant over the considered wavelength range. Results show that the nano-sized spirals can convert the optical power in a full width at half maximum (FWHM) of more than 20 THz because of its frequency-independent optical behavior.

The percent efficiency $\eta_e$ (%) = $\eta_e \times 100$ exhibited by the Seebeck antennas for each single frequency is shown in Fig. 4(c). By considering that the efficiency does not change with the intensity of the incoming wave, we can estimate the DC power $P^{tot}$ and the total harvesting efficiency $\eta^{tot}$ a single nanoantenna is expected to show under the solar irradiance [16,17]. These two quantities were estimated by using:

$$P^{tot} = \int_{\lambda_{start}}^{\lambda_{stop}} \eta_e \times P(\lambda,T) \times A_{eff} d\lambda \quad (5) \quad \text{and} \quad \eta^{tot} = \frac{\int_{\lambda_{start}}^{\lambda_{stop}} P(\lambda,T) \times \eta_e d\lambda}{\int_{\lambda_{start}}^{\lambda_{stop}} P(\lambda,T) d\lambda} \quad (6)$$

where $\lambda^{start}$ and $\lambda^{stop}$ are the starting and stopping wavelength of the mid-infrared band, and $P(\lambda,T)$ [W/m$^2$/nm] is the solar irradiance spectrum [16]. The DC power was estimated to be 1.7 aW and 4 aW, for the square and Archimedean spiral, respectively, while the total harvesting percent efficiency $\eta^{tot}$ (%) was $4.8 \times 10^{-7}$ and $3.4 \times 10^{-6}$ %.

On the other hand, both types of spiral thermocouples show a percent efficiency around $10^{-6}$ and $10^{-7}$. In order to better understand this value, we compare the efficiency $\eta_e$ with the maximum efficiency of thermocouples to convert the thermal energy into dc power $\eta_{th}$ (Fig. 5) and described by the equation (7) [23]:

$$\eta_{th} = \eta_C \frac{\sqrt{1+Z T_m} - 1}{\sqrt{1+Z T_m} + T_C/T_H} \quad (7)$$

where $\eta_C$ denotes the Carnot efficiency $1 - T_C/T_H$ which places a fundamental limit in all thermal engines, $T_m$ the average temperature between the hot and cold spots and $Z$ the thermoelectric figure-of-merit of thermocouple defined as [23,40],

$$Z_{AB} = (S_A - S_B)^2 / (\sqrt{\rho_A \kappa_A} + \sqrt{\rho_B \kappa_B}) \quad (8).$$

where $\kappa$ is the thermal conductivity coefficient and $\rho$ is the electric resistivity, being the subscripts $A$ and $B$ meaning the two materials involved in the thermocouple. These results show that the potential efficiency of the thermocouples at room temperature $\eta_{th}$ is greater than the conversion efficiency $\eta_e$ by a $10^2$ factor, revealing that most of the optical energy is dissipated by other mechanism as the heat absorbed by the substrate or the energy re-irradiated by antennas.

For the Seebeck nanoantennas to achieve the conversion efficiency of thermocouples, different strategies could be implemented such as suspending the device on air above its substrate (e.g. by using free-standing architecture).

In summary, the performance of thermoelectric nano-antennas to convert the optical power of infrared wavelengths into dc power was evaluated by using thermal numerical simulations, showing that devices represent an alternative way to harvest free-propagating optical energy (by exploiting the optical energy naturally dissipated as heat). The response of devices can be increased by reducing the effective thermal conductivity of the substrate. This can be achieved by suspending the device on air above its substrate. Moreover, engineering of large phase-arrays of nano-antennas acting as series thermocouples arrays can be implemented to increase responsivity of devices.

Acknowledgements

Support from CONACYT-Mexico under postdoctoral grant CV-45809, FONDECYT-Chile under project 3120059, Project ENE2009-013430 from the Spanish Ministerio de Innovación, Project "Centro Mexicano de Innovación en Energía Solar" from Fondo Sectorial CONACYT-Secretaría de Energía-Sustentabilidad Energética", and La Region Lorraine (France), is gratefully acknowledged.

Figures Captions:

FIG. 1. Schematic representation of the proposed Seebeck nanoantennas: (a) a square spiral; and (b) an Archimedean spiral. These geometries consist of two symmetrical arms made of dissimilar metals, what creates a thermocouple. For each arm the cross section is 200 nm width × 100 nm thickness. Substrates consist of 1.2 μm thick $SiO_2$ layers.

FIG. 2. DC equivalent circuit model of a Seebeck nanoantenna.

FIG. 3. (a) Temperature map of spirals Seebeck nanoantennas taken from a plane 50 nm below its surface. (b) Temperature profile all along the arms of the structures. The simulations were performed for two different polarization states: right-handed (RHCP) and left handed (LHCP) circular polarization of incident light at 10.6 μm.

FIG. 4. Simulations results of spiral Seebeck nanoantennas: (a) generated open voltage; (b) low-power dc signal supplied to the load resistance; (c) optical-to-electrical conversion efficiency. Blue squares refer to the square antenna, red triangles refer to the Archimedean antenna.

FIG. 5. Efficiency of thermocouples as a function of the incident frequency.

Figures:

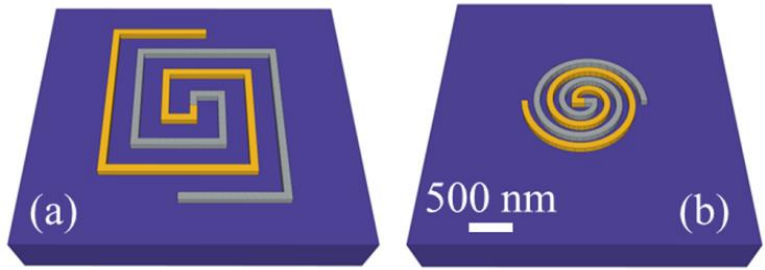

E. Briones *et al.* FIG. 1.

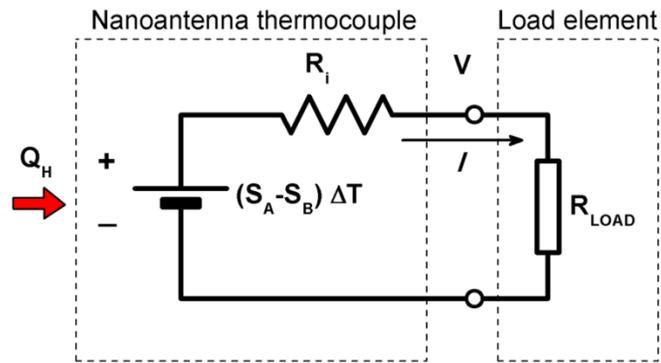

E. Briones *et al.* FIG. 2.

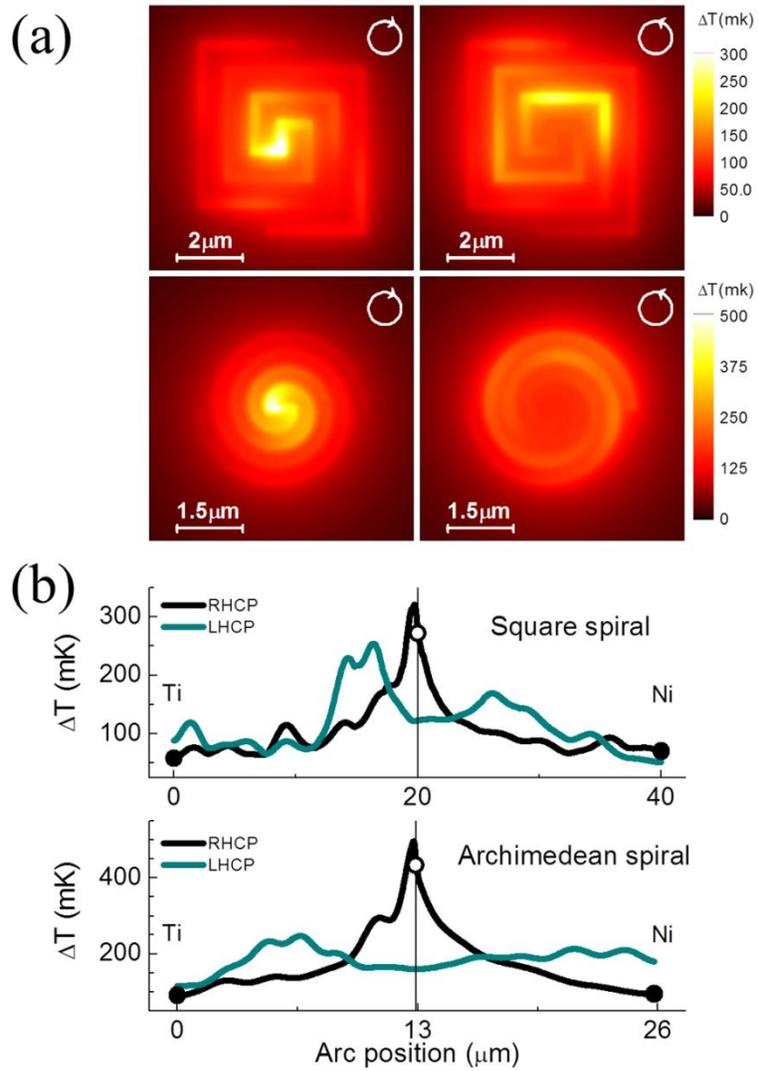

E. Briones *et al.* FIG. 3.

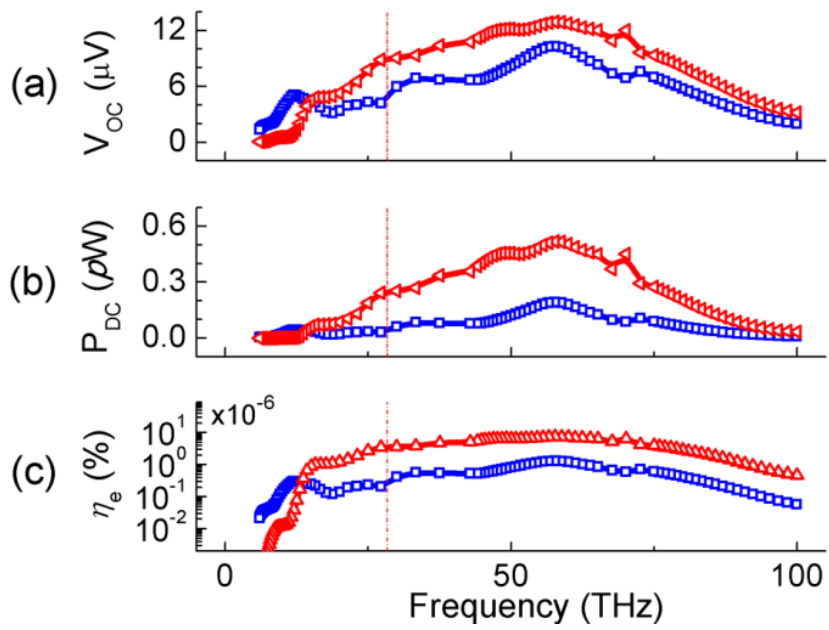

E. Briones *et al.* FIG. 4

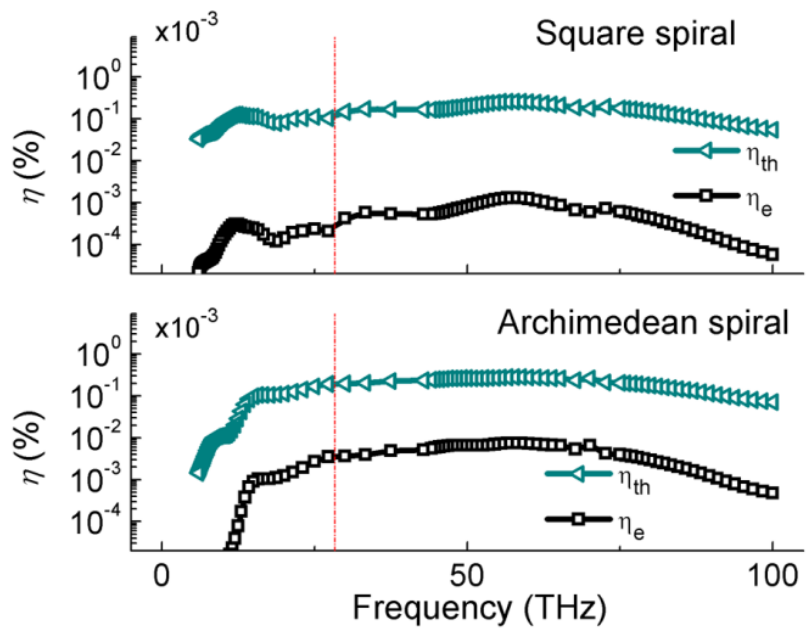

E. Briones *et al.* FIG. 5.

16